\documentstyle[preprint,pre,aps]{revtex}

\begin{document}
\draft
\title{
Period Doubling in Four-Dimensional Volume-Preserving Maps
}
\author{Sang-Yoon Kim \cite{byline}}
\address{
Department of Physics\\ Kangwon National University\\
Chunchon, Kangwon-Do 200-701, Korea
}
\maketitle
\begin{abstract}
We numerically study the scaling behavior of period doublings at the
zero-coupling critical point in a four-dimensional volume-preserving map
consisting of two coupled area-preserving maps.
In order to see the fine structure of period doublings, we extend the
simple one-term scaling law to a two-term scaling law. Thus we find a new
scaling factor $\delta_3$ $(=1.8505\dots)$ associated with scaling of the
coupling parameter, in addition to the previously known scaling factors
$\delta_1$ $(=-8.7210\dots)$ and $\delta_2$ $(=-4.4038\dots)$.
These numerical results confirm the renormalization results
reported by Mao and Greene [Phys. Rev. A {\bf 35}, 3911 (1987)].
\end{abstract}
%
%

\narrowtext

Universal scaling behavior of period doubling has been found in
area-preserving maps
\cite{Benettin,Collet,Greene,Bountis,Helleman,Widom,MacKay}.
As a nonlinearity parameter is varied, an initially stable periodic orbit may
lose its stability and give rise to the birth of a stable period-doubled
orbit. An infinite sequence of such bifurcations accumulates at a finite
parameter value and exhibits a universal limiting behavior. However these
limiting scaling behaviors are different from those
for the one-dimensional dissipative case
\cite{Feigenbaum}.

An interesting question is whether the scaling results of
area-preserving maps carry over higher-dimensional volume-preserving maps.
Thus period doubling in four-dimensional (4D) volume-preserving
maps has been much studied in recent years
\cite{MacKay,Janssen2,Mao1,Mao2,Kim1,Mao3}.
It has been found in Refs.~\cite{Mao2,Kim1,Mao3} that the critical scaling
behaviors of period doublings for two symmetrically coupled area-preserving
maps are much richer than those for the uncoupled area-preserving case.
There exist an infinite number of critical points in the space of the
nonlinearity and coupling parameters. It has been numerically found in
\cite{Mao2,Kim1} that the critical behaviors at those
critical points are characterized by two scaling factors, $\delta_1$ and
$\delta_2$. The value of $\delta_1$ associated with scaling of the
nonlinearity parameter is always the same as that of the
scaling factor $\delta$ $(=8.721\dots)$ for the area-preserving maps. However
the values of $\delta_2$ associated with scaling of the coupling parameter
vary depending on the type of bifurcation routes to the critical points.

The numerical results \cite{Mao2,Kim1} agree
well with an approximate analytic renormalization results obtained by
Mao and Greene \cite{Mao3}, except for the zero-coupling case in which
the two area-preserving maps become uncoupled.
Using an approximate renormalization method including truncation, they
found three relevant eigenvalues, $\delta_1=8.9474$,
$\delta_2=-4.4510$ and $\delta_3=1.8762$ for the zero-coupling case
\cite{Mao4}. However  they believed that
the third one $\delta_3$ is an artifact of the truncation, because
only two relevant eigenvalues $\delta_1$ and $\delta_2$ could be
indentified with the scaling factors numerically found.

In this paper we numerically study the critical behavior at the
zero-coupling point in two symmetrically coupled area-preserving maps
and resolve the inconsistency between the numerical
results on the scaling of the coupling parameter and the approximate
renormalization results for the zero-coupling
case. In order to see the fine structure of period doublings, we extend the
simple one-term scaling law to a two-term scaling law.
Thus we find a new
scaling factor $\delta_3=1.8505\dots$ associated with coupling, in
addition to the previously known coupling scaling factor $\delta_2
=-4.4038\dots$~.
The numerical values of $\delta_2$ and $\delta_3$ are close to the
renormalization results of the relevant coupling eigenvalues
$\delta_2$ and $\delta_3$.
Consequently the fixed map governing the critical behavior at the
zero-coupling point has two relevant coupling eigenvalues $\delta_2$ and
$\delta_3$ associated with coupling perturbations, unlike the cases of
other critical points.

Consider a 4D volume-preserving map consisting of two symmetrically coupled
area-preserving H${\acute {\rm e}}$non maps with a periodic boundary
condition,
\begin{equation}
T:\left\{
        \begin{array}{l}
         {\vec{\bf x}}(t+1) = - {\vec{\bf y}}(t) + {\vec{\bf F}}
         ({\vec{\bf x}}(t)), \\
         {\vec{\bf y}}(t+1) = {\vec{\bf x}}(t),
         \end{array}
   \right.
\label{eq:CM1}
\end{equation}
where ${\vec{\bf x}} = (x_1,x_2)$, ${\vec{\bf y}} = (y_1,y_2)$,
${\vec{\bf F}}({\vec{\bf x}}) = (F_1({\vec{\bf x}}),F_2({\vec{\bf x}}))$.
Here $z_m(t) \equiv (x_m(t),y_m(t))$ $(m=1,2)$ is the state vector of the
$m$th element at a discrete time $t$, and the periodic boundary condition
imposes $z_m=z_{m+2}$ for all $m$.
The $m$th component
$F_m({\vec{\bf x}})$ of the vector-valued function ${\vec{\bf F}}
({\vec{\bf x}})$ is given by
\begin{equation}
F_m({\vec{\bf x}}) = F(x_m,x_{m+1}) \nonumber \\
        = f(x_m) + g(x_m,x_{m+1}),
\end{equation}
where $f$ is the nonlinear function of the uncoupled H${\acute {\rm e}}$non's
quadratic map \cite{Henon}, i.e.,
\begin{equation}
f(x) = 1- ax^2,
\label{eq:NF}
\end{equation}
and $g({x_1},{x_2})$ is a coupling
function obeying a condition
\begin{equation}
g(x,x) = 0\;\;{\rm for\;\;any}\;\;x.
\label{eq:CC}
\end{equation}

The two-coupled map (\ref{eq:CM1}) is called a symmetric map \cite{Mao2,Kim1}
because it has an exchange symmetry such that
\begin{equation}
{\sigma}^{-1} T {\sigma} (\vec{\bf z}) = T (\vec{\bf z})\;\;\;{\rm for\;\;all}
\;\;\vec{\bf z},
\label{eq:ES}
\end{equation}
where $\vec{\bf z} = (z_1,z_2)$, $\sigma$ is an exchange operator acting on
$\vec{\bf z}$ such that
$\sigma {\vec{\bf z}} = (z_2,z_1)$, and $\sigma^{-1}$ is its
inverse.
The set of all fixed points of $\sigma$ is a two-dimensional (2D) subspace
of the 4D state space, on which $x_1=x_2$ and $y_1=y_2.$
It follows from Eq.~(\ref{eq:ES}) that the exchange operator $\sigma$
commutes with the symmetric map $T$, i.e., $\sigma T = T \sigma$.
Hence the 2D subspace becomes invariant under $T$.
An orbit is called an in-phase orbit if it lies on the 2D invariant subspace,
i.e., it satisfies
\begin{equation}
x_1(t) = x_2(t) \equiv x(t), \;\;
y_1(t) = y_2(t) \equiv y(t)\;\; {\rm for \;\;all}\;t.
\end{equation}
Otherwise it is called an out-of-phase orbit.
Here we study only in-phase orbits. They can be easily found from the
uncoupled H${\acute {\rm e}}$non map since the coupling
function $g$ obeys the condition (\ref{eq:CC}).

The Jacobian matrix $DT$ of the two-coupled map $T$ is:
\begin{equation}
DT = \left( \begin{array}{cc}
              D{\vec{\bf F}} & \;\; -I \\
               I & \;\; 0
            \end{array}
     \right),
\end{equation}
where $D{\vec{\bf F}}$ is the Jacobian matrix of the function
${\vec{\bf F}}({\vec{\bf x}})$,
$I$ is the $2 \times 2$ identity matrix, and 0 is the $2 \times 2$ null
matrix.
Since $Det(DT)=1$, the map $T$ is a 4D
volume-preserving map. Furthermore, if $D{\vec{\bf F}}$ is a symmetric matrix,
i.e., $D{\vec{\bf F}}^t=D{\vec{\bf F}}$ ($t$ denotes transpose), then the map
$T$ is a symplectic map because its Jacobian matrix satisfies the relation
$DT^t\;J\;DT = J$ \cite{Howard}, where
\begin{equation}
J = \left( \begin{array}{cc}
              0 & \;\; I \\
               -I & \;\; 0
            \end{array}
     \right).
\end{equation}

Stability analysis of an orbit in coupled maps is
conveniently carried out by Fourier-transforming with respect to the
discrete space $\{m\}$ \cite{Kapral}. Consider an orbit
$\{ {\vec{\bf x}} (t) \} \equiv \{ {x_m}(t)\; ; \;m=1,2 \}$ of the
two-coupled map $T$ expressed in the form of second-order difference
equations,
\begin{eqnarray}
T: {x_m}(t+1) + {x_m}(t-1) &=& {F_m}({\vec{\bf x}}(t)) \nonumber \\
 &=& f(x_m(t))+g(x_m(t),x_{m+1}(t)), \;\;m=1,2.
\label{eq:CM2}
\end{eqnarray}
The discrete spatial Fourier transform of the orbit is:
\begin{equation}
{\cal F}[{x_m(t)}] \equiv {\frac{1}{2}} {\sum_{m=1}^{2}}
{e^{-{\pi}imj}} {x_m}(t) = {\xi}_j(t),\;\;\;j=0,1.
\label{eq:FT}
\end{equation}
The wavelength of a mode with index $j$ is $\frac {2}{j}$.

To determine the stability of an in-phase orbit [$x_1(t) =x_2(t)
\equiv x(t)$ for all $t$],
we consider an infinitesimal perturbation $\{ {\delta}x_m(t) \}$
to the in-phase orbit, i.e.,
$x_m(t)=x(t)+{\delta}x_m(t)$ for $m=1,2$.
Linearizing the two-coupled map
(\ref{eq:CM2}) at the in-phase orbit, we have
\begin{equation}
{\delta}x_m(t+1) + {\delta}x_m(t-1)=f'(x(t))\;{\delta}x_m(t)+{\sum_{l=1}^{N}}
{G_l}(x(t))\; {\delta}x_{l+m-1}(t),
\label{eq:LE}
\end{equation}
where
\begin{equation}
f'(x)={\frac{df}{dx}},\;\;
\;{G_l}(x) =\left. { \frac{\partial g({\vec{\bf x}})}
{\partial x_l} } \right |_{x_1=x_2=x} .
\end{equation}
Hereafter we will call the functions $G_l$'s ``reduced'' coupling functions
of $g({\vec{\bf x}})$.

Let ${\delta {\xi}_j}(t)$ be the Fourier transform of {$\delta x_m(t)$},
i.e.,
\begin{equation}
\delta \xi_j (t)=
{\cal F}[{\delta x_m(t)}] = {\frac{1}{2}} {\sum_{m=1}^{2}}
{e^{-{\pi}imj}} {\delta x_m}(t),\;\;j=0,1.
\end{equation}
Then the Fourier transform of Eq.~(\ref{eq:LE}) becomes:
\begin{equation}
\delta {\xi}_j(t+1) +\delta {\xi}_j(t-1) =
[f'(x(t))+{\sum_{l=1}^{2}} {G_l}(x(t))
{e^{\pi i(l-1)j}}]\;{\delta {\xi}_j}(t),\;\;j = 0,1.
\label{eq:LM1}
\end{equation}
This equation can be also put into the following form:
\begin{equation}
\left(
\begin{array}{l}
\delta{\xi}_j(t+1) \\
\delta{\xi}_j(t)
\end{array}
\right)
= L_j(t)
\left(
\begin{array}{l}
\delta{\xi}_j(t) \\
\delta{\xi}_j(t-1)
\end{array}
\right),\;\;j=0,1,
\label{eq:LM2}
\end{equation}
where
\begin{equation}
L_j(t)=
\left( \begin{array}{cc}
f'(x(t))+{\displaystyle{\sum_{l=1}^{2}}} {G_l}(x(t)) {e^{\pi i(l-1)j}} &
\;\; -1 \\
1 &\;\; 0
        \end{array}
\right).
\label{eq:JML}
\end{equation}
Note that the determinant of $L_j$ is one, i.e., $Det(L_j)=1$.

It follows from the condition (\ref{eq:CC}) that the reduced coupling
functions satisfy
\begin{equation}
{\sum_{l=1}^{2}} G_l(x) = 0.
\end{equation}
Hence there exists only one independent reduced coupling function $G(x)$
such that
\begin{equation}
{G_2}(x)=G(x),\,\,\,{G_1}(x)=-G(x).
\end{equation}
Substituting $G_l$'s into the $(1,1)$ entry of the matrix $L_j(t)$, we
have:
\begin{equation}
{\sum_{l=1}^{2}} G_l(x(t)) e^{\pi i(l-1)j} =
\left \{ \begin{array}{c}
          0\;\;{\rm for}\;\; j=0, \\
          -2 G(x(t))\;\;{\rm for}\;\; j =1
         \end{array}
\right.
\label{eq:SE}
\end{equation}
Hence the matrices $L_j$'s of Eq.~(\ref{eq:JML}) are real matrices.

Stability of an in-phase orbit of period $q$ is
determined by iterating Eq.~(\ref{eq:LM2}) $q$ times:
\begin{equation}
\left(
\begin{array}{l}
\delta{\xi}_j(t+q) \\
\delta{\xi}_j(t+q-1)
\end{array}
\right)
= M_j
\left(
\begin{array}{l}
\delta{\xi}_j(t) \\
\delta{\xi}_j(t-1)
\end{array}
\right),\;\;j=0,1,
\end{equation}
where
\begin{equation}
M_j= {\prod_{k=t}^{t+q-1}} L_j(k).
\end{equation}
That is, the stability of each mode with index $j$ is
determined by the $2 \times 2$ matrix $M_j$.
Since $Det(M_j)=1$, each matrix $M_j$ has a reciprocal
pair of eigenvalues, $\lambda_j$ and $\lambda_j^{-1}$.
These eigenvalues are called the stability multipliers of the mode with
index $j$.
Associate with a pair of multipliers $(\lambda_j,\lambda_j^{-1})$
a stability index,
\begin{equation}
\rho_j = \lambda_j + \lambda_j^{-1},\;\;j=0,1,
\end{equation}
which is just the trace of $M_j$, i.e., $\rho_j=Tr(M_j)$.
Since $M_j$ is a real matrix, $\rho_j$ is always real.
Note that the matrix $M_0$ for the case of the $j=0$ mode is just the
Jacobian matrix of the uncoupled H${\acute {\rm e}}$non map.
Hence the coupling affects only the stability index $\rho_1$ of
the $j=1$ mode.

It follows from the reality of $\rho_j$ that the reciprocal pair of
eigenvalues of $M_j$ lies either on the
unit circle, or on the real line in the complex plane, i.e., they are
a complex conjugate pair on the unit circle, or a reciprocal pair of reals.
Each mode with index $j$ is stable if and only if its
stability index $\rho_j$ is real with $|\rho_j| \leq 2$, i.e., its
stability multipliers are a pair of complex conjugate numbers of modulus
unity.
A period-doubling (tangent) bifurcation occurs when
the stability index $\rho_j$ decreases (increases) through $-2$ (2), i.e.,
two eigenvalues coalesce
at $\lambda_j = -1$ (1) and split along the negative (positive) real axis.

An in-phase orbit is stable only when all its modes are stable. Hence its
stable region in the space of the nonlinearity and coupling parameters is
bounded by four bifurcation lines associated with tangent and
period-doubling bifurcations of both modes (i.e., those curves determined
by the equations $\rho_j = \pm 2$ for $j=0,1$).
When the stability index $\rho_0$ decreases
through $-2$, the in-phase orbit loses its stability via in-phase
period-doubling bifurcation and gives rise to the birth of the period-doubled
in-phase orbit. Here we are interested in scaling behaviors of such in-phase
period-doubling bifurcations.

As an example we consider a linearly-coupled case in which the coupling
function is
\begin{equation}
g(x_1,x_2)= {c \over 2} (x_2 - x_1).
\label{eq:CF}
\end{equation}
Here $c$ is a coupling parameter. Figure \ref{figure1} shows the stability
diagram of in-phase orbits with period $q=1,2,$ and $4$ \cite{Rem1}.
As previously observed in \cite{Mao2,Kim1}, each ``mother'' stability
region bifurcates into two ``daughter''
stability regions successively in the parameter plane; hereafter we call
the direction of the left (right) branch of the two daughter
stability regions $L$ $(R)$ direction.
Consequently the stability region of the in-phase orbit with period $q=2^n$
$(n=0,1,2,\dots)$ consists of $2^n$ branches. Each branch can be represented
by its address $[a_0,\dots,a_n]$, which is a sequence of symbols $L$ and
$R$ such that $a_0 =L$ and $a_i =$ $L$ or $R$ for $i \geq 1$.

An infinite sequence of connected stablity branches (with increasing period)
is called a bifurcation ``route'' \cite{Mao2,Kim1}. Each bifurcation route is
also represented by an infinite sequence of symbols $L$ and $R$.
A ``self-similar'' bifurcation ``path'' in a bifurcation route is formed
by following a sequence of parameters $(a_n,c_n)$, at which the in-phase
orbit of level $n$ (period $2^n$) has some given stability indices
$(\rho_0, \rho_1)$ (e.g., $\rho_0=-2$ and $\rho_1=2$) \cite{Mao2,Kim1}.
All bifurcation paths within a bifurcation route  converge to an
accumulation point $(a^*,c^*)$, where the value of $a^*$ is always the
same as that of the accumulation point for the area-preserving case
(i.e., $a^*=4.136\,166\,803\,904 \dots$), but
the value of $c^*$ varies depending on the bifurcation routes.
Thus each bifurcation route ends at a
critical point $(a^*, c^*)$ in the parameter plane.

It has been numerically found that scaling behaviors near a critical point
are characterized by two scaling factors, $\delta_1$ and $\delta_2$
\cite{Mao2,Kim1}. The value of $\delta_1$ associated with scaling of the
nonlinearity parameter is always the same as that of the scaling factor
$\delta$ $(=8.721\dots)$ for the area-preserving case. However the values
of $\delta_2$ associated with scaling of the coupling parameter vary
depending on the type of bifurcation routes. These
numerical results agree well with analytic renormalization results
\cite{Mao3}, except for the case of one specific
bifurcation route, called the $E$ route. The address of the $E$ route is
$[(L,R,)^\infty]$ $(\equiv [L,R,L,R,\dots])$ and it ends at the
zero-coupling critical point $(a^*,0)$.

Using an approximate renormalization method including truncation, Mao and
Greene \cite{Mao4} obtained three relevant eigenvalues, $\delta_1 = 8.9474$,
$\delta_2 = -4.4510$, and $\delta_3 = 1.8762$ for the zero-coupling case;
hereafter the two eigenvalues $\delta_2$ and $\delta_3$ associated with
coupling will be called the coupling eigenvalues (CE's).
The two eigenvalues $\delta_1$ and $\delta_2$ are close to the numerical
results of the nonlinearity-parameter scaling factor $\delta_1(=8.721\dots)$
and the coupling-parameter scaling factor $\delta_2(=-4.403\dots)$ for the
$E$ route.
However they believed that the second relevant CE $\delta_3$ is an
artifact of the truncation,
because it could not be identified with
anything obtained by a direct numerical method.

In order to resolve the inconsistency between the numerical results and
the renormalization results for the zero-coupling case, we numerically
reexamine the scaling behavior associated with coupling.
Extending the simple one-term scaling law to a two-term scaling law, we
find a new scaling factor
$\delta_3 = 1.8505\dots$ associated with coupling in addition to the
previously found coupling scaling factor $\delta_2=-4.4038\dots$, as will
be seen below. The values of these two coupling scaling factors are close
to the renormalization results of the relevant CE's $\delta_2$ and
$\delta_3$.

We follow the in-phase orbits of period $2^n$ up to level $n=14$
in the $E$ route and
obtain a self-similar sequence of parameters $(a_n,c_n)$, at which
the pair of stability indices, $(\rho_{0,n},\rho_{1,n})$, of the orbit
of level $n$ is $(-2,2)$.
The scalar sequences $\{ a_n \}$ and $\{ c_n \}$ converge  geometrically
to their limit values, $a^*$ and $0$, respectively.
In order to see their convergence, define
$\displaystyle{ \delta_n \equiv {\Delta a_{n+1} \over \Delta a_n}}$ and
$\displaystyle{ \mu_n \equiv {\Delta c_{n+1} \over \Delta c_n}},$
where $\Delta a_n = a_n - a_{n-1}$ and
$\Delta c_n = c_n - c_{n-1}$.
Then they converge to their limit values $\delta$ and $\mu$ as $n \rightarrow
\infty$, respectively. Hence the two sequences $\{ \Delta a_n \}$ and
$\{ \Delta c_n \}$ obey one-term scaling laws asymptotically:
\begin{equation}
\Delta a_n = C^{(a)} \delta^{-n},\;\;\;
\Delta c_n = C^{(c)} \mu^{-n}\;\;\;{\rm for\;large\;}n,
\label{eq:OTSL}
\end{equation}
where  $C^{(a)}$ and $C^{(c)}$ are some constants.

The two sequences $\{ \delta_n \}$ and $\{ \mu_n \}$  are shown in Table
\ref{table1}.
The second column shows rapid convergence of $\delta_n$ to its limit value
$\delta$ $(=8.721\dots)$, which is close to the renormalization result of
the first relevant eigenvalue (i.e., $\delta_1=8.9474$).
The sequence $\{ \mu_n \}$ also seems to converge
to $\mu = -4.403\dots$ (see the third column). However its convergence is
not as fast as that for the case of the sequence $\{ \delta_n \}$. This is
because two relevant CE's $\delta_2$ and $\delta_3$ are involved in the
scaling of the sequence $\{ \mu_n \}$.

Taking into account the effect of the second relevant CE $\delta_3$ on the
scaling of the sequence $\{ \Delta c_n \}$,
we extend the simple one-term scaling law (\ref{eq:OTSL}) to a two-term
scaling law:
\begin{equation}
\Delta c_n = C_1 \mu_{1}^{-n} + C_2 \mu_{2}^{-n} \;\;\;{\rm for\;large\;}n,
\label{eq:TTSL1}
\end{equation}
where $| \mu_1 | > | \mu_2 |$.
This is a kind of multiple scaling law \cite{Mao5,Reick}.
Eq.~(\ref{eq:TTSL1}) gives
\begin{equation}
\Delta c_n = t_1 \Delta c_{n+1} -t_2 \Delta c_{n+2},
\label{eq:RE}
\end{equation}
where $t_1 = \mu_1 + \mu_2$ and $t_2 = \mu_1 \mu_2$.
Then $\mu_1$ and $\mu_2$ are solutions of the following quadratic equation,
\begin{equation}
\mu^2 - t_1 \mu + t_2 =0.
\label{eq:QE}
\end{equation}
To evaluate $\mu_1$ and $\mu_2$, we first obtain $t_1$ and $t_2$ from
$\Delta a_n$'s using Eq.~(\ref{eq:RE}):
\begin{equation}
t_1 = { {\Delta c_n \Delta c_{n+1} - \Delta c_{n-1} \Delta c_{n+2}}
\over {\Delta c_{n+1}^2 - \Delta c_n \Delta c_{n+2}} },\;\;\;
t_2 = { {\Delta c_n^2 - \Delta c_{n+1} \Delta c_{n-1}}
\over {\Delta c_{n+1}^2 - \Delta c_n \Delta c_{n+2}} }.
\label{eq:T1T2}
\end{equation}
Note that Eqs.~(\ref{eq:TTSL1})-(\ref{eq:T1T2}) hold only for large $n$.
In fact the values of $t_i$'s and $\mu_i$'s $(i=1,2)$ depend on the
level $n$. Therefore we explicitly denote $t_i$'s and $\mu_i$'s by
$t_{i,n}$'s and $\mu_{i,n}$'s, respectively. Then each of them converges
to a constant as $n \rightarrow \infty$:
\begin{equation}
\lim_{n \rightarrow \infty} t_{i,n} = t_i, \;\;\;
\lim_{n \rightarrow \infty} \mu_{i,n} = \mu_i,\;\;i=1,2.
\end{equation}

Three sequences $\{ \mu_{1,n} \}$, $\{ \mu_{2,n} \}$, and
$\displaystyle{ \{ {\mu_{1,n}^2 / \mu_{2,n}} \} }$ are shown in
Table \ref{table2}.
The second column shows rapid convergence of $\mu_{1,n}$ to its limit
values $\mu_1$ $(=-4.403\,897\,805)$. Comparing this sequence with
the sequence $\{ \mu_n \}$ in Table I, we find that the finite sequence
$\{ \Delta c_n \}$ obeys the two-term scaling law better than the one-term
scaling law. That is, the accuracy for the scaling factor $\mu_1$ obtained
from the two-term scaling law is much higher than that obtained from the
one-term scaling law. The value of this scaling factor $\mu_1$ is close to
the renormalization result of the first relevant CE (i.e.,
$\delta_2$ $=-4.4510$).
 From the third and fourth columns, we also find that the second scaling
factor $\mu_2$ is given by a product of two relevant CE's $\delta_2$ and
$\delta_3$,
\begin{equation}
\mu_2 = {\delta_2^2 \over \delta_3},
\end{equation}
where $\delta_2=\mu_1$ and $\delta_3=1.850\,65$~.
It has been known that every scaling factor in the multiple-scaling
expansion of a parameter is expressed by a product of the eigenvalues of a
linearized renormalization operator \cite{Mao5,Reick}.
Note that the value of $\delta_3$ is close to the renormalization result
of the second relevant CE (i.e., $\delta_3=1.8762$).

We now study the coupling effect on the stability index $\rho_{1,n}$
of the $j=1$ mode of the in-phase orbit of period $2^n$ near the
zero-coupling critical point $(a^* , 0)$.
Figure \ref{figure2} shows three plots of $\rho _{1,n} (a^*,c)$ versus
$c$ for $n=4,5,$ and $6$. For $c=0$,
$\rho_{1,n}$ converges to a constant $\rho_1^*$ $(=-2.543\,510\,20\dots)$,
called the critical stability index \cite{Kim1}, as $n \rightarrow \infty$.
However, when $c$ is non-zero $\rho_{1,n}$ diverges as $n \rightarrow
\infty$, i.e., its slope $S_n$
$\displaystyle{ \left.
(\equiv  {{\partial \rho_{1,n}} \over {\partial c}} \right|_{(a^* ,0)})
}$
at the zero-coupling critical point
diverges as $n \rightarrow \infty$.

The sequence $\{ S_n \}$ obeys a two-term scaling law,
\begin{equation}
S_n = D_1 \nu_1 ^n + D_2 \nu_2 ^n\;\;\;{\rm for\;large\;}n,
\label{eq:TTSL2}
\end{equation}
where $|\nu_1| > |\nu_2|$.
This equation gives
\begin{equation}
S_{n+2} = r_1 S_{n+1} - r_2 S_{n},
\end{equation}
where $r_1 = \nu_1 + \nu_2$ and $r_2 = \nu_1 \nu_2$.
As in the scaling for the coupling parameter, we first obtain
$r_1$ and $r_2$ of level $n$ from $S_n$'s:
\begin{equation}
r_{1,n} = {  {S_{n+1} S_{n} - S_{n+2} S_{n-1}}
\over {S_{n}^2 - S_{n+1} S_{n-1}} },\;\;\;
r_{2,n} = { { S_{n+1}^2 - S_{n} S_{n+2}}
\over {S_{n}^2 - S_{n+1} S_{n-1}} }.
\end{equation}
Then the scaling factors $\nu_{1,n}$ and $\nu_{2,n}$ of level $n$ are given
by the roots of the quadratic equation, $\nu_n^2 - {r_{1,n}} {\nu_n} +
{r_{2,n}} =0$. They are listed in Table \ref{table3} and converge to
constants $\nu_1$ $(=-4.403\,897\,805\,09)$ and $\nu_2$ $(=1.850\,535)$
as $n \rightarrow \infty$, whose accuracies are higher than those of the
coupling-parameter scaling factors.
Note that the values of $\nu_1$ and $\nu_2$ are
also close to the renormalization results of the two relevant CE's $\delta_2$
and $\delta_3$.

We have also studied several other coupling cases with the coupling function,
$g(x_1 , x_2) = {\displaystyle {c \over 2}} (x_2^n - x_1^n) $
($n$ is a positive integer).
In all cases studied $(n=2,3,4,5)$, the scaling factors of both the coupling
parameter $c$ and the slope of the stability index $\rho_1$ are found to
be the same as those for the above linearly-coupled case $(n=1)$ within
numerical accuracy. Hence universality also seems to be well obeyed.

\acknowledgments

I thank Professor S. Y. Lee for a critical reading of the manuscript.

%
%
\begin{table}
\caption{ Scaling factors $\delta_n$ and $\mu_n$ in the one-term
          scaling for the nonlinearity and coupling parameters are
          listed.
         }
\begin{tabular}{ccc}
$n$ & $\delta_n$ & $\mu_n$ \\
\tableline
6 & 8.721\,086\,300\,39 &  -4.399\,51 \\
7 & 8.721\,096\,265\,95 &  -4.405\,75 \\
8 & 8.721\,097\,056\,68 &  -4.403\,12 \\
9 & 8.721\,097\,188\,39 & -4.404\,22 \\
10& 8.721\,097\,198\,70 & -4.403\,76  \\
11& 8.721\,097\,200\,44 & -4.403\,96 \\
12& 8.721\,097\,200\,58 & -4.403\,87 \\
13& 8.721\,097\,200\,60 & -4.403\,91 \\
\end{tabular}
\label{table1}
\end{table}

\begin{table}
\caption{ Scaling factors $\mu_{1,n}$ and $\mu_{2,n}$ in the two-term
          scaling for the coupling parameter are shown in the second
          and third columns, respectively. A product of them,
          $\displaystyle{ {\mu_{1,n}^2 \over \mu_{2,n}} }$, is shown in the
          fourth column.}
\begin{tabular}{cccc}
$n$ & $\mu_{1,n}$ & $\mu_{2,n}$ &
$\displaystyle{ {\mu_{1,n}^2 \over \mu_{2,n}} }$ \\
\tableline
5 & -4.403\,908\,128 & 10.437\,4 & 1.858\,17 \\
6 & -4.403\,899\,694 & 10.465\,9 & 1.853\,09 \\
7 & -4.403\,898\,736 & 10.458\,2 & 1.854\,46 \\
8 & -4.403\,897\,867 & 10.474\,8 & 1.851\,52 \\
9 & -4.403\,897\,847 & 10.473\,9 & 1.851\,68 \\
10& -4.403\,897\,806 & 10.478\,4 & 1.850\,89 \\
11& -4.403\,897\,807 & 10.478\,6 & 1.850\,85 \\
12& -4.403\,897\,805 & 10.479\,7 & 1.850\,65 \\
\end{tabular}
\label{table2}
\end{table}

\begin{table}
\caption{ Scaling factors $\nu_{1,n}$ and $\nu_{2,n}$ in the two-term
          scaling for the slope of the stability index of the $j=1$ mode
          are shown.}
\begin{tabular}{ccc}
$n$ & $\nu_{1,n}$ & $\nu_{2,n}$ \\
\tableline
5 & -4.403\,898\,453\,59 &  1.851\,433\,5 \\
6 & -4.403\,897\,730\,29 &  1.850\,782\,6 \\
7 & -4.403\,897\,813\,85 &  1.850\,603\,6 \\
8 & -4.403\,897\,804\,07 &  1.850\,553\,8 \\
9 & -4.403\,897\,805\,21 &  1.850\,540\,0 \\
10& -4.403\,897\,805\,07 &  1.850\,536\,1 \\
11& -4.403\,897\,805\,09 &  1.850\,535\,0 \\
12& -4.403\,897\,805\,09 &  1.850\,534\,9 \\
\end{tabular}
\label{table3}
\end{table}
%
%
\begin{figure}
\caption{Stability diagram of in-phase orbits for the linearly-coupled case.
        The horizontal (non-horizontal) solid and dashed lines denote the
        period-doubling and tangent bifurcation lines of the $j=0$ $(1)$
        mode, respectively. For other details see the text.}
\label{figure1}
\end{figure}
\begin{figure}
\caption{Plots of the stability index $\rho_{1,n} (a^*,c)$ versus $c$
        for $n=4,5,6$. ~~~~~~~~~~~~~~~~~~~~~~~~~~~}
\label{figure2}
\end{figure}
%
%

\begin{references}
\bibitem{Rem1} Only some upper part of its
whole stable region is shown for the case of the period-1 orbit
in order to make the stable regions for the cases of higher
periods $(q=2,4)$ clearer.
\end{references}
\end{document}